\documentclass[
reprint,
 amsmath,amssymb,
 aps,
]{revtex4-2}

\usepackage{graphicx}
\usepackage{dcolumn}
\usepackage{bm}

\begin{document}

\author{A. Kucherik}
\affiliation{Department of Physics and Applied Mathematics, Stoletov Vladimir State University, 600000 Gor’kii street, Vladimir, Russia}

\author{R. R. Hartmann}
\affiliation{Physics Department, De La Salle University,  2401 Taft Avenue, 0922 Manila, Philippines}

\author{A. V. Povolotskiy}
\affiliation{Institute of Chemistry, St.~Petersburg State University, 198504 Ulianovskaya street, St.~Petersburg, Russia}

\author{A. Osipov}
\author{V. Samyshkin}
\affiliation{Department of Physics and Applied Mathematics, Stoletov Vladimir State University, 600000 Gor’kii street, Vladimir, Russia}

\author{M. E. Portnoi}
\email{M.E.Portnoi@exeter.ac.uk}
\affiliation{Physics and Astronomy, University of Exeter, Stocker Road, Exeter EX4 4QL, United Kingdom}

\title{Polarization-sensitive photoluminescence from aligned carbon chains terminated by gold clusters}

\keywords{narbon chains, photoluminescence spectra, nanoparticles, plasmons }

\begin{abstract}
We synthesize a thin film composed of long carbyne chains terminated by gold clusters and study its optical properties. The presence of gold particles stabilizes longer chains and leads to their alignment. We show that the gold clusters also act as a source of electron doping thus changing the intensity of photoluminescence from quadratic dependence on the pumping intensity without gold to linear with gold. We also observe that the excitation of the film at the gold plasmon frequency causes the blue shift of photoluminescence and estimate on the basis of this effect the minimum length of the carbyne chains. The high degree of alignment of the gold-terminated carbyne chains results in strongly anisotropic light absorption characterized by a distinctive cosine dependence on the angle between the carbyne molecule and polarization plane of the excitation. This paves the way for a new class of ultimately-thin polarization sensitive emitters to be used in future integrated quantum photonics devices.
\end{abstract}

\maketitle

After several decades of unsuccessful experimental attempts and controversies\cite{eisler2005polyynes}, the synthesis of long chains of carbon atoms, with alternating single and triple bonds between them, known as linear acetylenic carbon or carbyne\cite{shi2016confined}, has now become a reality\cite{bianco2018carbon}. Due to its ultimate one-dimensional (1D) nature, carbyne exhibits extremely anisotropic optical properties. In addition, unlike most carbon nanotubes, which suffer from poor quantum efficiency due to dark excitons arising from their multivalley band structure, carbyne is a direct bandgap material characterized by strong photoluminescence in the visible range\cite{xiao2017molecular}. 

A recent theoretical study has shown that individual carbyne chains should exhibit strong light absorption and emission when excited by light that is polarized along the chain\cite{hartmann2021terahertz}. This motivates optical experiments with films of highly aligned carbynes excited by linearly polarized light with the ultimate goal of creating ultra-thin polarization sensitive elements of optoelectronic devices.

Carbyne is a highly reactive and unstable material, which makes its synthesis challenging. However, finite length carbyne chains can now be synthesized using a variety of techniques ranging from laser ablation\cite{pan2015carbyne,kutrovskaya2020excitonic}, to chemical vapor deposition, electrochemical reduction and low-temperature pyrolysis\cite{khanna2017formation}. While a true carbyne crystal is composed of an infinitely long chain of carbon atoms\cite{zhang2020review}, finite length carbyne chains can still display distinctive optical and electronic properties that render them promising candidates for a diverse range of optoelectronic applications\cite{yang2022synthesis}. Indeed, the length of the carbyne chain, and edge termination play an important role in determining its optical properties and behavior\cite{hartmann2021terahertz}.  

In this Letter we demonstrate that a thin film made of aligned carbyne chains terminated by gold nanoparticles is characterized by photoluminescence with the intensity dependent on the angle between the chain alignment direction and the polarization plane of the exciting light. A striking feature of the system is that not only do the gold nanoparticles stabilize and align the carbyne chains, but they also dope the chains with electrons, leading to a dramatic enhancement of the photoluminescence intensity compared to intrinsic carbyne. The photoluminescence intensity from gold-terminated chains is linearly dependent on the pump intensity, in contrast to the quadratic intensity dependence of a much weaker luminescence from undoped chains. In addition to the usual red-shifted luminescence mentioned above, we observe blue-shifted luminescence when the excitation frequency is close to the plasmon resonance of the gold clusters. This effect requires long but finite length chains, and we provide an estimate of the minimum required length of these chains based on the experimentally observed blue-shift data.

\begin{figure}
    \centering
\includegraphics[width=\columnwidth]{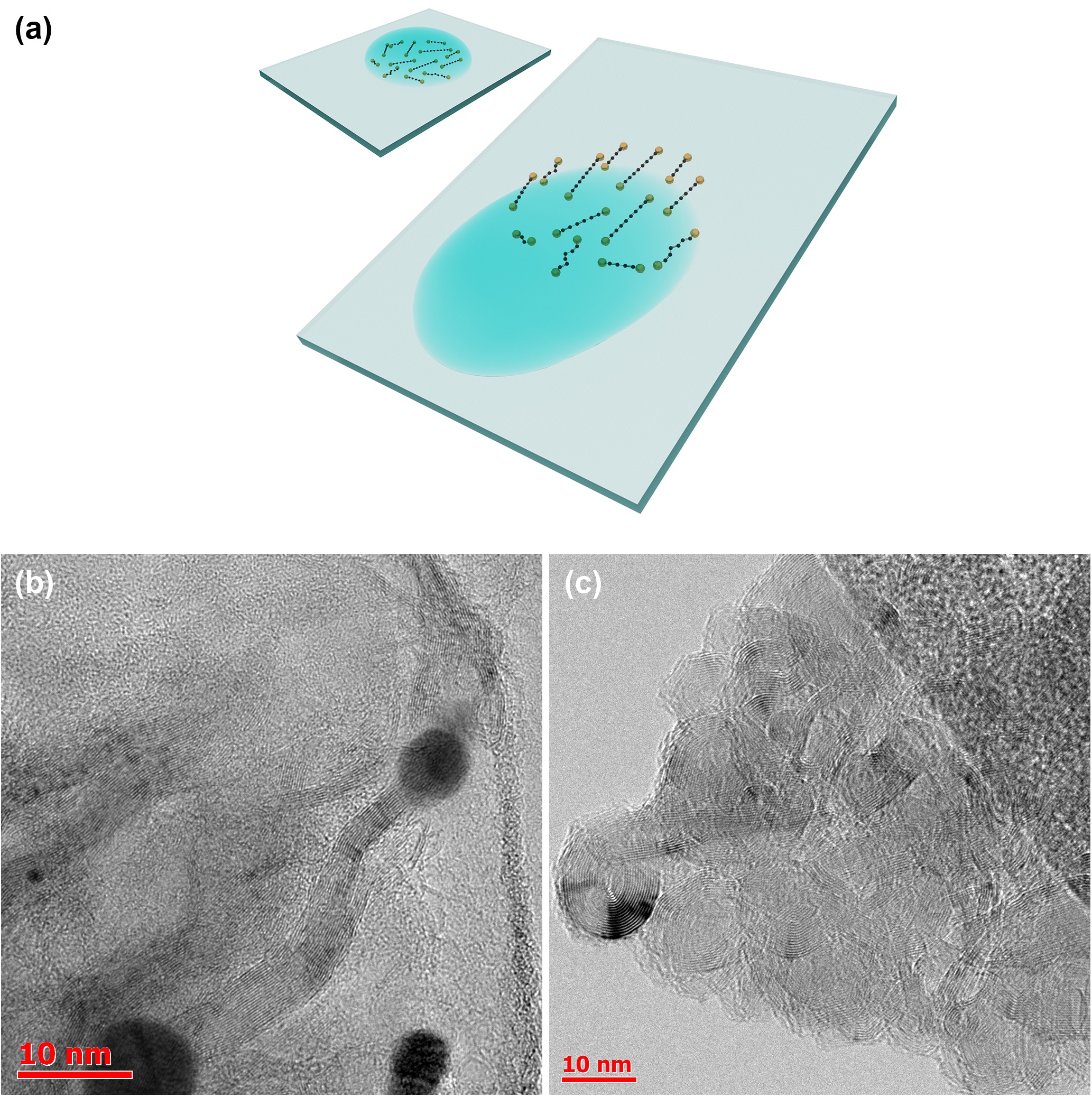}
    \caption{A sketch (a) depicting the process of obtaining aligned chains by tilting the substrate during the drying of the colloidal solution containing long gold-terminated carbynes. TEM-images of the irradiated colloidal system: (b) with gold particles; (c) only shungite carbon. The gold nanoparticles appear as dark regions in the image. 
}
    \label{fig1}
\end{figure}

To produce the ensembles of aligned carbyne chains stabilized by gold nanoparticles used in our photoluminescence experiments, we followed the methodology outlined in Ref.\citenum{kutrovskaya2020excitonic}. However, to increase the number of linear chains, we tripled the concentration of amorphous carbon. The colloidal solution was deposited onto quartz substrates, and dried at room temperature following the method presented in Fig.~\ref{fig1}~(a) which employs the effect of liquid-solid adhesion (pinning effect)\cite{hu2005analysis}. Fifteen seconds after deposition, the substrate, placed on a rotary table, was tilted up to 20 degrees. Upon tilting the substrate the droplet of colloidal solution, which is approximately 1 cm in diameter, slowly flows down towards the lower edge of the substrate. After that, the upper boundary of the droplet was studied for the effect of the formation of extended linear structures. It can be seen from the TEM image shown in Fig.~\ref{fig1}~(b) that these chains can extend to hundreds of atoms in length. In the absence of the stabilizing influence of gold nanoparticles, the laser fragmentation method also results in the formation of carbon chains\cite{marabotti2022pulsed}. However, without gold they fold into more stable structures such as spirals and coils (see Fig.~\ref{fig1}~(c)), or fragment into randomly-aligned shorter molecules.

\begin{figure}[ht]
    \centering
    \includegraphics[width=\columnwidth]{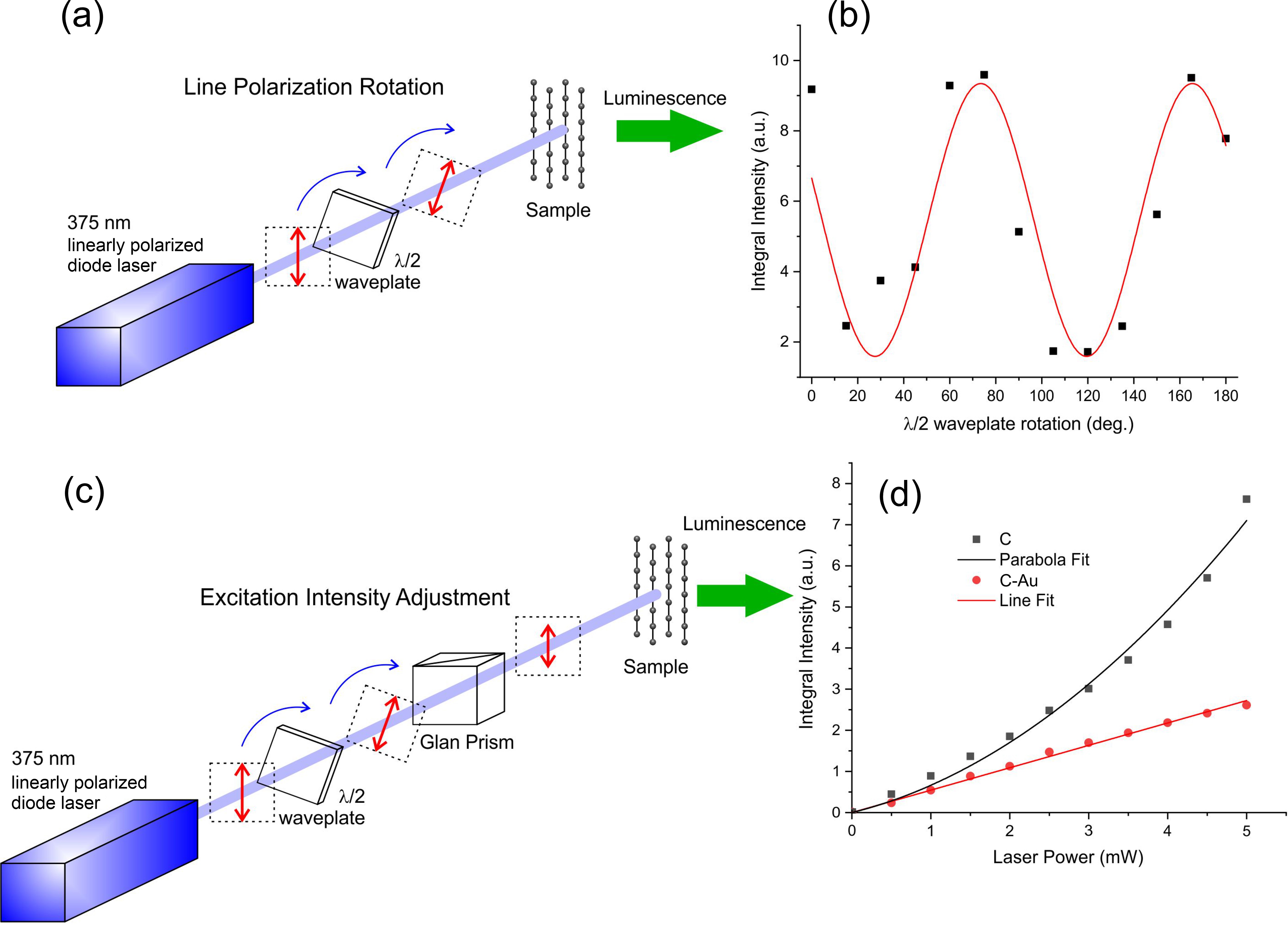}
    \caption{
    Experimental setup for photoluminescence measurements and their results: (a) a sketch showing the rotation of the polarization plane of excitation, with (b) the corresponding photoluminescence dependence on rotation angle for gold-terminated chains; (c) a sketch of the setup to change the pump intensity at fixed polarization angle, with (d) the resulting measurements for both the irradiated colloidal system containing (red markers) and not containing gold clusters (black markers). The results for pure carbon samples have been multiplied by a factor of hundred to enable visual comparison. 
}
    \label{fig3}
\end{figure}

The experimental setup for measuring the intensity dependence of photoluminescence on excitation polarization angle is shown in Fig.~\ref{fig3}~(a). For optical excitation, we used a 375 nm continuous wave laser (Coherent CUBE), and the resulting photoluminescence was detected using a Fluorolog-3 spectrofluorometer (Horiba). The angle between the polarization plane of the optical excitation and the alignment axis of carbyne chains is changed using a rotating half-wavelength plate. From Fig.~\ref{fig3}~(b), it can be clearly seen that the sample exhibits highly anisotropic absorption of light, which manifests itself in the strong dependence of integral luminescence intensity on the polarization angle of the exciting light. Therefore the alignment of chains within the sample occurs at length scales that are significantly larger than the wavelengths used in the experiment. At the same time the thickness of the film is several orders of magnitude smaller than the wavelength, making this system a unique ultra-thin polarization-sensitive element with the potential use in nanoscale devices utilizing the polarization properties of light to be employed in quantum cryptography and computations. It should also be noted that the irradiated colloidal sample synthesized without gold nanoparticles shows no dependence of photoluminescence on the polarization angle.

\begin{figure}[ht]
    \centering
    \includegraphics[width=0.9\columnwidth]{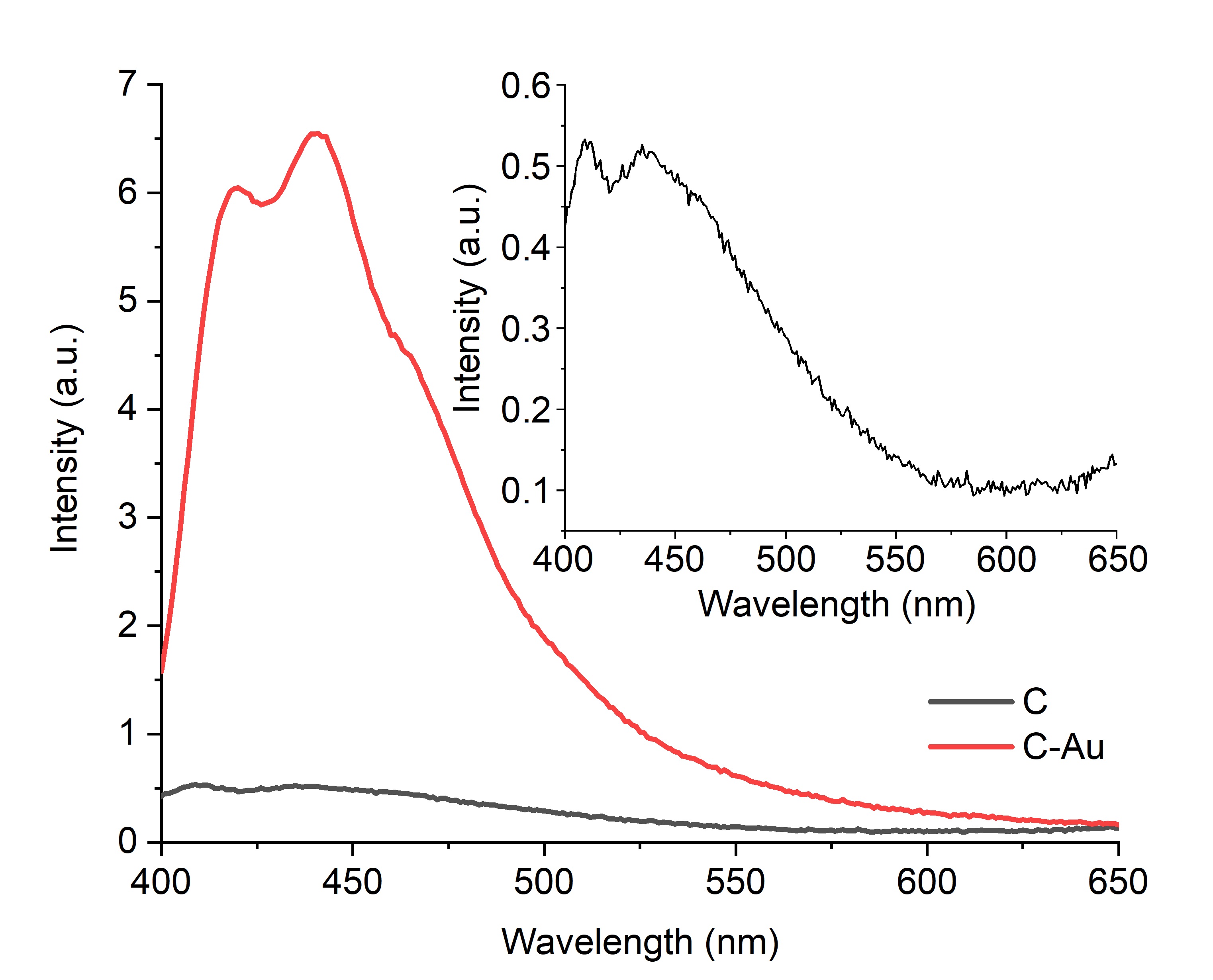}
    \caption{     
The photoluminescence spectra of the carbyne chains synthesized in the presence of gold nanoparticles (red line) and in the absence of gold nanoparticles (gray line) are shown. The laser excitation wavelength used was 375 nm. The inset shows the magnified photoluminescence spectrum of the gold-free sample.
    }
    \label{fig4}
\end{figure}

Another striking feature of carbyne terminated by gold, as shown in Fig.~\ref{fig4}, is the significant enhancement of its photoluminescence intensity, which is an order of magnitude greater than that from the samples without gold. Notably, the observed photoluminescence spectra are typical for linear carbon chains decorated by gold nanoparticles\cite{yang2020visible,chen2020fluorescent}. In addition to significant emission enhancement, the intensity of the photoluminescence of gold-terminated chains is linearly dependent on the excitation intensity in contrast to the quadratic intensity dependence for the gold-free samples. We performed photoluminescence intensity measurements using the setup shown in Fig.~\ref{fig3}~(c), keeping the polarization plane of the light beam fixed 
for samples synthesized with and without gold clusters. The results of these measurements are shown in Fig.~\ref{fig3}~(d), from which the difference in the excitation intensity dependence of integral photoluminescence for the two types of samples is clearly seen.

\begin{figure}[ht]
    \centering
    \includegraphics[width=\columnwidth]{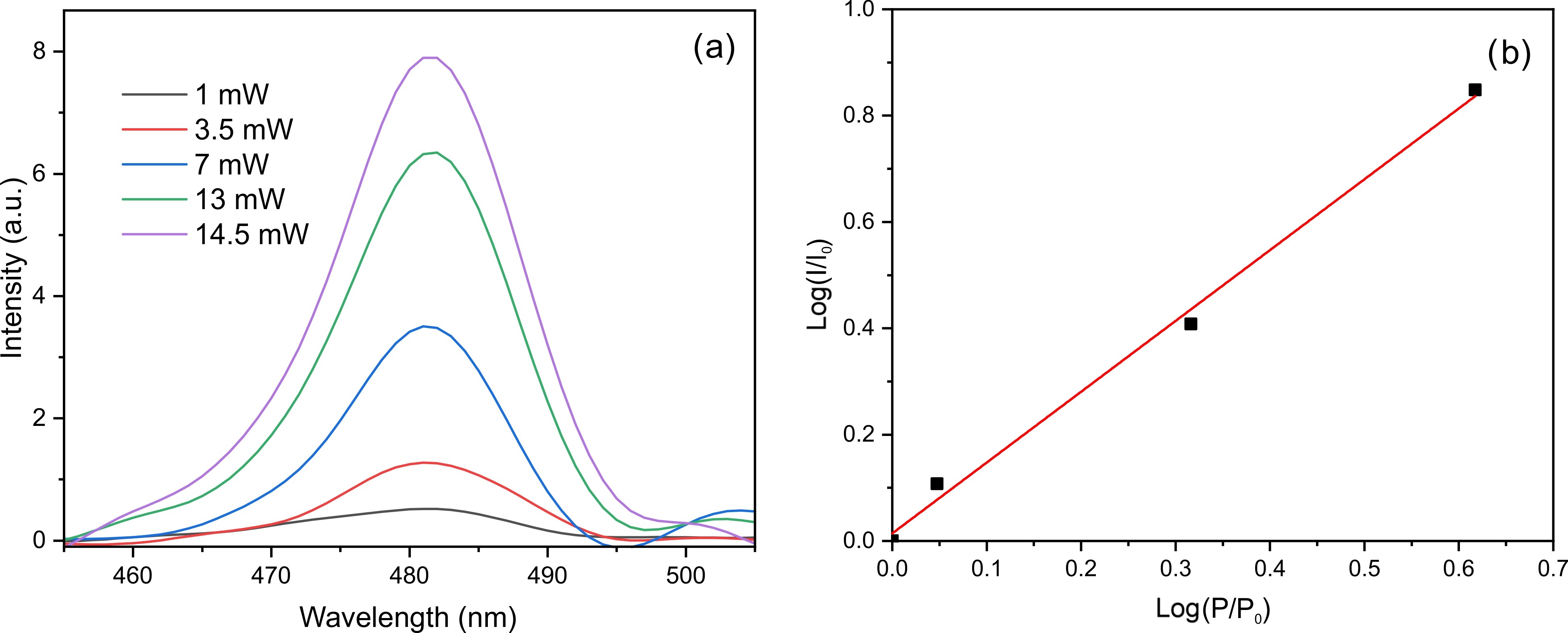}
    \caption{The observed luminescence spectrum (a) for the sample excited at the gold plasmon resonance frequency for several values of pumping intensity. Panel (b) shows the double logarithm plot of the integral luminescence intensity on the excitation power. Here $I_0$ and $P_0$ are the maximum luminescence intensity and pumping power values, respectively.
}
    \label{fig5}
\end{figure}

We also excited the gold-enriched samples using a laser with a frequency corresponding to the plasmon resonance of gold. The substrate contribution to the photoluminescence was subtracted, revealing a broad peak centered around 480~nm, see Fig.~\ref{fig5}~(a). The photoluminescence spectra are blue-shifted by 30-80~nm from the pump laser wavelength of 532~nm. The spectral width of the employed monochromator, which is 5~nm, is well below the possible experimental error. The data was collected from a wide sample area of approximately 5~$\mathrm{mm}^{2}$. The pump power dependence of the integral photoluminescence intensity is shown in Fig.~\ref{fig5}~(b) using a double logarithmic scale. This dependence is linear, with a slope of approximately 1.3, indicating a non-linear dependence of integral photoluminescence on the excitation power.  

The ability of carbyne to act as a polarizer, along with the giant enhancement of its photoluminescence due to electron doping by gold nanoparticles, as well as the dependence of photoluminescence intensity on pump strength, can be easily interpreted within a model of dipole optical transitions in a simple finite-length chain described by the nearest-neighbour tight-binding Hamiltonian with two alternating hopping integrals.

The probability of an optical transition between energy levels in a linear carbon chain is given by 
\begin{equation}
    W_{i\to f}\propto
    I\left|\langle f|\boldsymbol{r}\cdot\hat{\boldsymbol{e}}|i
    \rangle\right|^{2}
    \mathcal{L}\left(E_{i}-E_{f}-h\nu\right)
    F_i F_f ,
    \label{eq:transition}
\end{equation}
where $I$ is the intensity of the excitation, $i$ and $f$ are the initial and final states, $E_i$ and $E_f$ are
their corresponding energies, $\boldsymbol{r}$ is the position operator, $\nu$ the photon frequency, $\boldsymbol{e}$ the light polarization vector, and $F_i$ ($F_f$) is the probability that the initial (final) state is occupied (unoccupied). The bell-shaped function $\mathcal{L}\left(E_{i}-E_{f}-h\nu\right)$ results from the broadening by various relaxation processes of the Delta function entering the Fermi's golden rule. We believe that the dominating process broadening the optical transition peaks is the escape of photoexcited electrons into gold for the higher states, and the injection of electrons from gold into the lower states of the chains emptied by light (escape of holes from the chains into gold). 

The exact form of the matrix element of transition $\left|\langle f|\boldsymbol{r}\cdot\hat{\boldsymbol{e}}|i\rangle\right|$ for finite carbon chains is given in Ref.\citenum{hartmann2021terahertz}. The main feature of the system is the alternating parity of consecutive energy levels, whereas dipole transitions only occur between states of different parity. Since in a linear molecule the direction of $\boldsymbol{r}$ is fixed and the optical selection rules depend on $\boldsymbol{r}\cdot\hat{\boldsymbol{e}}$,  rotating the polarization plane of the incident light modifies the intensity of photoluminescence by a factor of $\cos^2(\phi)$, where $\phi$ is the angle between the polarization vector of the incident light, and the carbyne axis. In our experimental setup, the half-wave plate rotates the polarization angle by $2\theta$, where $\theta$ is the angle between the fast axis of the polarizer and the polarization plane of the incident light. It can be seen from Fig.~\ref{fig3}~(b), that the intensity of the photoluminescence of the irradiated sample containing gold nanoparticles obeys the relation $I\propto\cos^2(2\theta + \phi)$, whereas the irradiated sample containing no gold nanoparticles shows no such dependence.

\begin{figure}[ht]
    \centering
\includegraphics[width=\columnwidth]{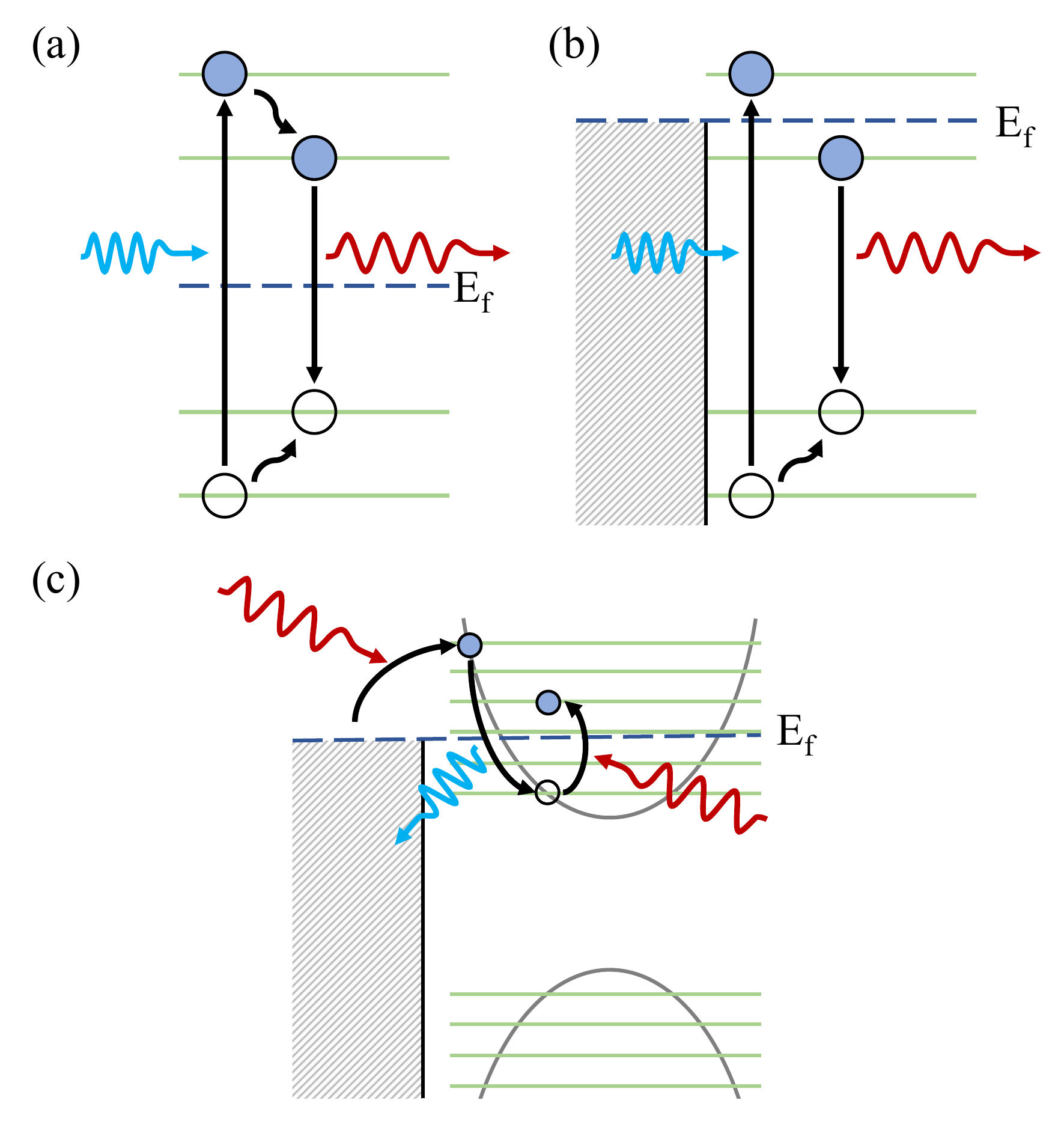}
    \caption{The scheme of optical transitions for (a) undoped carbyne chains; (b) carbyne chains doped by electrons supplied from gold clusters; and (c) gold-terminated carbyne chains excited at the gold plasmon frequency.
    }
    \label{fig:transitions}
\end{figure}

Equation~(\ref{eq:transition}) also allows us to explain the observed striking difference in the photoluminescence data with and without gold clusters mentioned above. Let us first consider an undoped carbon chain whose ends are not terminated by gold nanoparticles. For simplicity we shall consider a short molecule, with discrete energy levels, but the explanation given herein is also valid for the continuum model to be employed for extra long chains. As schematically depicted in Fig.~\ref{fig:transitions}~(a), illuminating the sample optically will lead to the promotion of electrons to an energy level above the Fermi level, and the creation of holes below the Fermi level. The photoexcited electrons and holes can relax to the LUMO and HOMO levels via the emission of a photon. 
As can be seen from Eq.~\ref{eq:transition}, the probability of a LUMO-HOMO transition is proportional to the probability of a photoexcited electron occupying the LUMO level, product the probability of a photoexcited holes occupying the HOMO level, i.e., $F_{i}F_{j}=F_{i}^2$. Therefore, increasing the power of illumination will increase the observed luminescence of an ensemble of carbon chains quadratically.

Now let us consider the case of gold-terminated carbon chains. During the synthesis process the gold nanoparticles may acquire a negative charge\cite{scanlon2015charging}, resulting in the Fermi level being higher than the LUMO level of an undoped carbon chain. In this situation, electrons will transfer from the gold into the carbon chains. Upon high frequency illumination, the electrons are promoted to an energy level above the Fermi level. As shown in Fig.~\ref{fig:transitions}~(b), the photoexcited holes float up emitting photons to the level corresponding to the LUMO of the undoped system. In stark contrast to the system without gold this level is completely occupied by electrons, therefore the luminescence intensity in the weak pumping regime is proportional only to the probability that a photoexcited hole is present. Therefore, the observed luminescence in an ensemble of gold terminated carbon chains is linearly dependant on the pump laser power. 

It should also be noted that the same model also explains the weak blue-shifted photoluminescence of gold-terminated chains upon excitation close to the gold plasmon frequency shown in Fig.~\ref{fig5}. First, it is important to note that a distinction exists between an infinite carbyne crystal and the finite length carbon chains studied in this research. While only direct dipole transitions between valence and conduction bands are allowed in an infinite crystal, transitions between discreet energy levels of opposite parity are allowed in finite chains. These transitions correspond to the forbidden indirect transitions in the continuum model within the conduction band. Let us now consider a chain long enough such that energy levels are closely spaced, as shown in Fig.~\ref{fig:transitions}~(c). The HOMO-LUMO gap is analogous to the band gap of a bulk semiconductor, and similarly to a metal-semiconductor interface, a Schottky barrier forms between the gold nanoparticle and the carbon chains\cite{kutrovskaya2022light}. As illustrated in Fig.~\ref{fig:transitions}~(c), an electron can be optically excited from the gold nanoparticle, to a carbyne level above the Fermi level. The same excitation frequency can also result in a carbyne electron being promoted from the below the Fermi level to above it. The hot electron injected from the gold can then recombine with the photoexcited hole, resulting in blue-shifted photoluminescence. As was mentioned above, for this process to occur the finite-length chain should be long enough so that there are several ``conduction band'' levels (unoccupied molecular orbitals) within the energy range corresponding to the excitation frequency. In fact, a simple analysis involving the parity selection rules shows that there should be at least three carbyne energy levels within the excitation photon energy. This requirement for the inter-level transition to occur at the gold plasmon energy ($h\nu\approx2.33$~eV) yields the estimate for the minimum number of carbon atoms in the chain: $12\left|\beta\right|/N<h\nu$, where $\left|\beta\right|\approx3.5$~eV is the long bond hoping integral (using the notation adopted from Ref.\citenum{hartmann2021terahertz}). This estimate provides $N=18$ as the minimum number of atoms for the blue shift to be observed.

It can be seen from Fig.~\ref{fig5}~(b) that the integral dependence of the blue-shifted luminescence is superlinear with a power of approximately $4/3$, which is weaker than the quadratic dependence of red-shifted luminescence of gold-free samples since the number of photocreated holes in the chain is linearly proportional the excitation power, whereas the number of hot electrons injected from gold due to plasmon excitation has a more complex sublinear dependence on the pumping power\cite{kutrovskaya2022light}.

In conclusion, we synthesized and studied the optical properties of ultimately thin carbon chains. When these chains are terminated by gold clusters they become highly aligned with absorption strongly dependent on the angle between the chains and the polarization plane of the excitation. The presence of gold clusters also results in the doping of the chains by free electrons leading to the photoluminescence intensity to be linearly dependent on the excitation power. The luminescence from gold terminated chains is much stronger than that from the pure carbon structures which have quadratic dependence of photoluminescence intensity on the pumping power. When gold-terminated chains are illuminated by a laser with a frequency close to the plasma frequency of gold clusters there is a weak blue-shifted luminescence observed. All these effects can be described by a model of finite-length chain inter-level transitions broaden by the exchange of carriers between the carbon chains and gold clusters. The presence of blue-shifted luminescence allows to estimate the minimum length of the chains. The strong polarization dependence of the luminescence intensity of gold-terminated carbon chain arrays makes them promising candidates for nanoscale logic elements in emerging light-controlled quantum devices.

\begin{acknowledgments}
This work was supported by RSF-grant 23-12-20004. R.R.H. and M.E.P were supported by the EU H2020-MSCA-RISE projects TERASSE (H2020-823878) and DiSeTCom (H2020-823728). R.R.H. acknowledges financial support from URCO (15 F 2TAY21 - 3TAY22). A.V.P. acknowledges support from St.~Petersburg State University  (project $\#94031307$). The optical measurements were conducted using the equipment from the Centre for Optical and Laser Materials Research at St.~Petersburg State University Research Park.
\end{acknowledgments}


\end{document}